\documentstyle[prl,aps,epsf]{revtex}
\draft
\begin{document}
\twocolumn[\hsize\textwidth\columnwidth\hsize\csname @twocolumnfalse\endcsname

\title{Critical Currents and Vortex States at
Fractional Matching Fields in Superconductors with
Periodic Pinning}  
\author{Charles Reichhardt}
\address{Department of Physics, University of California, Davis, California
95616.}

\author{Niels Gr{\o}nbech-Jensen}
\address{Department of Applied Science, University of California,
Davis, California 95616.\\
NERSC, Lawrence Berkeley National Laboratory, Berkeley, California 94720.}

\date{\today}
\maketitle
\begin{abstract}
We study vortex states and dynamics 
in 2D superconductors with periodic pinning at fractional  
sub-matching fields 
using numerical simulations. 
For square pinning arrays we show that ordered vortex states 
form at 
$1/1, 1/2$, and $1/4$ filling fractions 
while only partially ordered states form at
other filling fractions, such as $1/3$ and $1/5$,   
in agreement with recent imaging experiments. For triangular  
pinning arrays we observe matching effects 
at filling fractions of 
$1/1, 6/7, 2/3, 1/3, 1/4, 1/6$, and $1/7$.
For both square and triangular pinning arrays we also find that, 
for certain sub-matching fillings, 
vortex configurations depend on pinning strength. For 
weak pinning, ordering in which a portion 
of the vortices are positioned between pinning sites can occur.  
Depinning of the vortices at the 
matching fields, where the vortices are ordered, is elastic while
at the incommensurate fields the motion is plastic.
At the incommensurate fields, as the applied driving force is increased,
there can be a transition to elastic flow where the vortices
move along the pinning sites in 1D channels and a reordering transition
to a triangular or distorted triangular lattice. We also discuss the
current-voltage curves and how they relate to the vortex ordering at
commensurate and incommensurate fields. 
\end{abstract}

\vskip2pc]
\narrowtext

\section{Introduction}

Vortex lattices interacting with quenched disorder represent an ideal system
in which to study various static and dynamic phases of elastic lattices
in disordered media, which can be found in a large class of condensed
matter systems. Pinning of vortices is also of practical importance
since superconductors are required to maintain high critical currents
for various potential technological applications. There have been considerable
efforts to enhance the pinning properties of vortices in superconductors
by introducing point, columnar, and splay defects using electron, proton
and heavy ion irradiation \cite{Blatter1}. 
A particularly promising technique has been 
to lithographically create nano-structured periodic defect arrays. 
Experiments with arrays of holes 
\cite{Fiory2,Metl3,Metlushk4,Look5,DeLong6,Puig7,Moshchalkov8,Wodenweber9,Harada10,Welp11,Field12,Bending13} and magnetic dots
\cite{Schuller14,Hoffman15,Jaccard16,Fasano17,Terentiev18,Shuller219}, 
as well as 
theory \cite{Jacard20} and numerical simulations \cite{Commensurate21}, have
shown that a variety of commensurability effects occur when the density
of vortices equals an integer multiple of the density of pinning sites
such that a well ordered vortex crystal can be stabilized and pinned.
Several experiments have provided evidence that for small pinning sites, 
beyond the first matching field, a portion of the vortices will sit at the
interstitial regions between the pinning sites \cite{DeLong6,Harada10}.  

Besides matching effects at integer filling fractions, commensurability
effects at non-integer and sub-matching fields have also been observed
directly with Lorentz microscopy \cite{Harada10} and indirectly with
magnetization and transport measurements \cite{Metlushk4,Look5,DeLong6}. 
Experiments by Metlushko {\it et al.} observed sub-matching effects down
to $1/16$ filling fractions \cite{Look5}. These same experiments did not
show any particular fractional matching for $B/B_{\phi} > 1.0$. In other
experiments, fractional matching has been seen at filling fractions of
$3/2$ and $5/2$ \cite{Welp11}. These fractional matching effects are
generally less pronounced than those of the integer matching fields. 

Recent scanning-Hall probe measurements have also directly imaged vortex
configurations at various sub-matching fields \cite{Field12,Bending13}.
These measurements find that some configurations are only partially
ordered while others, including the integer matching fields, are almost
completely ordered. Experiments by Field {\it et al.}, who image up to
5,000 vortices, have found that at the 1/2 matching field very large regions
of ordered vortex checkerboard states can be observed with well defined
domain walls separating different orientations
\cite{Field12}. At other fields, such as
$1/3$, only domains of diagonally ordered regions are observed, while at
$1/4$, $2/5$, and $1/5$ even smaller regions of order are observed.  
 
Previous simulations of vortices interacting with periodic pinning arrays
have focused on vortex matching at integer matching fields and have found
that different types of vortex crystals can be stabilized at different
matching fields in agreement with experiments \cite{Commensurate21}. 
The case of vortex matching at fractional fields for square and triangular
pinning arrays has not been addressed. Although commensurate vortex
configurations in Josephson-junction arrays \cite{Halsey22} and repulsive
particles on lattices \cite{Watson23} at fractional fillings have been
studied, these models do not allow the vortices to sit at interstitial  
regions between the pinning sites as is possible in experiments with
periodic pinning, magnetic dot, or hole arrays. For $B/B_{\phi} > 1$
it is clear that interstitial pinning of vortices is possible. In this
work we show that for certain pinning strengths and commensurabilities 
it is possible for a portion of the vortices to be positioned at interstitial 
sites even for $B/B_{\phi} < 1$, so that the vortices form a type of
ordering that would be absent for vortices in wire networks and
Josephson-junction arrays. Further, the dynamic phases are also strongly
affected by the fact that vortex flow can occur through the interstitial
regions. 

For {\it random pinning} sites, theory 
\cite{Koshelev24,Giamarchi25,Balents26},
simulations \cite{Koshelev24,Dominguez27,Nori28} and experiments
\cite{Higgins29,Pardo30,Troya31} find that an initially pinned disordered
vortex lattice can first flow plastically and then undergo a dynamic
transition from a disordered structure to an ordered moving vortex
lattice which can be a moving Bragg glass or moving smectic phase.
The vortices in the strongly driven phase flow in well defined
channels and a transverse depinning threshold can be present 
\cite{Giamarchi25}.
Similarly, in simulations with {\it periodic pinning} where the vortex
density exceeds the pinning density, $B/B_{\phi} > 1$, distinct kinds
of plastic flow and large transverse barriers may exist \cite{Periodic32}. 
Simulations of Josephson-junction arrays have also found plastic and
elastic vortex flow phases and transverse depinning transitions
\cite{Candia33,Pastoriza34}.

The dynamics of vortices in periodic pinning arrays with $B/B_{\phi} < 1$
has not previously been studied. For this case, some of the open questions
are {\it (i)} whether the vortex configurations in the pinned state are
the same as the moving vortex configuration; {\it (ii)} how this depends on
the filling fraction; and {\it (iii)} whether there can be plastic flow and how
it depends on commensurability. Unlike systems with random pinning, 
where the disorder comes from the quenched substrate, the disorder in
vortex systems with periodic pinning comes about from the vortex-vortex
interactions with the incommensurate fields forming a disordered vortex
lattice. At the commensurate fields, where the vortex lattice is ordered,
plastic flow may be absent even though the vortex lattice is very
strongly pinned. At strong driving forces it may be interesting to see if
the incommensurate vortex lattice can reorder to a moving crystal phase
in a similar manner as observed in systems with random pinning. It may also
be interesting to compare commensurate and incommensurate fields 
in strongly driven phases to see whether the vortex flow occurs in 1D
channels and if the flow is between or along the pinning rows. It may further
be valuable to correlate the microscopics of the static and dynamical
phases to macroscopic observables, such as critical currents,
current-voltage characteristics, and magnetization. 

In this work we examine the vortex configurations and dynamics for square
and triangular pinning arrays in 2D superconductors with logarithmically
interacting vortices using molecular dynamics simulations of pancake vortices.
For the square pinning array we find a series of sub-matching effects that
can be seen as peaks in the critical depinning force. The vortex
configurations at these fields agree well with those seen in recent imaging 
experiments \cite{Harada10,Field12,Bending13}. Certain sub-matching fields, such
as $1/3$ are not completely ordered, but are broken up into domains as also 
observed in experiments \cite{Field12}. Vortex ordering at $1/3$ filling
depends on the pinning strength with stronger pinning producing a partially
diagonally ordered arrangement. For weak disorder at $1/3$ filling we observe
a vortex ordering in which certain vortices sit in the interstitial regions
between the pinning sites so that the overall vortex lattice has a near
triangular ordering. We also find that the critical depinning force
increases linearly with pinning force strength at the commensurate fields
while for low pinning strengths the incommensurate field shows a 
nonlinear (sub-linear) increase due to vortex-vortex interactions
which crosses over to a linear increase.  

We find that at $B/B_{\phi} = 1, 1/2$, and $1/4$, where the vortex lattice
is ordered, the depinning process and subsequent flow is elastic with the
moving vortex lattice having the same symmetry as the pinned lattice. For
fields near strong commensurability, and for $B/B_{\phi} > 1$, we observe
a two-stage depinning process where vacancies or interstitial defects in
the commensurate vortex crystal depin first, as can be seen in velocity
versus driving force curves that are equivalent to experimental
$V(I)$ curves.

At the incommensurate fields, for $B/B_{\phi} > 0.35$, the vortex lattice
depins plastically with only a portion of the vortices moving and the
vortex motion occurring in winding paths in both the $x$ and $y$ directions. 
For these fields (vortex densities) at strong drives we observe
{\it reordering} to an elastic distorted triangular moving lattice where
the vortices move in {\it 1D channels} along the pinning rows. For
$B/B_{\phi} < 0.35$ the initial depinning is still plastic but is followed by a 
transition to a channel flow phase where the vortices move in 1D channels
along the pinning rows. In contrast to what occurs at higher fields,
the individual 1D channels at lower fields contain different numbers of
vortices, and the moving vortex lattice retains considerable positional
disorder. These reordered phases are only stable in the strongly driven
limit and we do not observe any hysteresis in the $V(I)$ curves or vortex
lattice order for either field range. 

For triangular pinning arrays we find a different set of matching filling
fractions than those found for square pinning. The effective pinning of the
lattice is enhanced and an ordered vortex lattice 
appears at $B/B_{\phi} = 1$, $6/7$,
$2/3$, $1/3$, $1/4$, $1/6$, and $1/7$. Further, there is no peak in the 
critical
depinning force at $1/2$. For strong pinning the vortex lattice at 
$B/B_{\phi} = 1/2$ is disordered and all the vortices are located at 
pinning sites, while for weak disorder the vortex lattice at $1/2$ 
has a square ordering in which half of the vortices are located at the
interstitial regions. For weak pinning the commensurability peaks in the
critical depinning force are enhanced and an additional peak at
$B/B_{\phi} = 1/2$ is observed. For both strong and weak pinning the
moving vortex lattice at $B/B_{\phi} = 1/2$ is square. For the triangular
pinning array we do not find any matching effects for $B/B_{\phi} > 1$ 
which is due to the fact that the interactions with interstitial vortices
cause a certain portion of the pinning sites to be left unoccupied.
We observe $V(I)$ characteristics and moving phases similar to the ones
found in the square pinning array.    

\section{Simulation}

We model logarithmically interacting pancake vortices in a 2D superconductor
interacting with square and triangular pinning arrays. The overdamped
equation of motion for a vortex $i$ is 
\begin{equation}
{\bf f}_{i} = \frac{d {\bf r}_{i}}{dt} = 
{\bf f}_{i}^{vv} + {\bf f}_{i}^{vp} + {\bf f}_{d} = {\bf v}_{i}
\; .
\end{equation}
The vortex-vortex interaction potential is $U_{v} = -\ln (r)$. The total
force force on vortex $i$ from the other vortices is
$f_{i}^{vv} = -\sum_{j\neq i}^{N_{v}}\nabla_i U_{v}(r_{ij})$. 
We impose periodic
boundary conditions and evaluate the periodic long-range logarithmic
interaction with an exact sum \cite{Jensen35}. Pinning is modeled as
attractive parabolic wells of radius $r_{p}$,
\begin{equation} 
f_{i}^{vp} = -\sum_{k=1}^{N_p}(f_{p}/r_{p})({\bf r}_{i}-{\bf r}_{k}^{(p)})
 \ \Theta ((r_{p} - |{\bf r}_{i}- {\bf r}_{k}^{(p)}|)/\lambda)
\end{equation}

\noindent
Here, we measure distance in units of the penetration depth
$\lambda$, ${\bf r}_{k}^{(p)}$ is the location of pinning
site $k$, $f_{p}$ is the maximum pinning force, and $\Theta$ is the
Heaviside step function 
The pinning is placed in a square or triangular array. 
The initial vortex positions are obtained by annealing from a high temperature
configuration, with the vortices in a molten state, and slowly cooling
to $T=0$. This is intended to mimic field cooled experiments performed
in \cite{Field12,Bending13}. We reduce the initial temperature to zero
in twenty steps and remain at each step for $1.5\times10^{5}$ MD steps,
each being $dt=0.01$ in normalized units. 
The vortex configurations are insensitive
to the waiting time for waiting times greater than $5\times10^{4}$.
We note that the true ground states may only be attainable for prohibitively
long waiting times; however, the agreement between our results and
experiments is encouraging. 

After the vortex configurations are obtained, the critical depinning force is 
determined by applying a slowly increasing, spatially uniform driving force,
$f_{d}$, which would correspond to a Lorentz force from an applied current. 
For each drive increment we measure the average vortex velocity in the
direction of drive, $V_{x} = \sum_{i}^{N_{v}}{\hat {\bf x}}\cdot {\bf v}_{i}$.  
The force, $f_{d}$, versus velocity, $V_{x}$, curve corresponds
experimentally to a voltage-current, $V(I)$, curve. The depinning force
is defined to be the force value at which $V_{x}$ reaches a value of $0.03$
times the ohmic (linear) response. 

\section{Fractional Matching For Square Pinning Array}

In Fig.~1(a,b) we show the critical depinning force, $f_{p}^{c}$, versus
the vortex density for a system with a pinning density of
$n_{p} = 1.36/\lambda^{2}$, $f_{p} = 0.9f_{0}$, and $r_{p} = 0.3\lambda$. 
In Fig.~1(a) we show $f_{p}^{c}$ for magnetic fields up to
$B/B_{\phi} = 2.25$, illustrating the matching effects at $B/B_{\phi} = 1$
and $2$ as well as clear matching effects at $1/2$ and $1/4$.
Figure~1(b) displays $f_{dp}^{c}$ for $B/B_{\phi} < 0.6$, where additional
sub-matching peaks are visible at $1/3$, $1/6$, $1/5$, and $1/8$, with the
$1/3$ matching field being the largest of these. We do not observe any
particular matching at $2/3$ or $4/5$. Figure~2(a) shows the vortex
positions (black circles) and pinning sites (open circles) for
$B/B_{\phi} = 1$, where the vortices form a square lattice with the pinning
sites. In Fig.~2(b), we show the vortex configuration at an incommensurate
field, $B/B_{\phi} = 0.64$, where a peak in the critical depinning force is
absent. Here, no apparent ordering in the vortex lattice can be observed.
We also find that the vortex lattice is in general disordered at the other
incommensurate fields. In Fig.~2(c), for $B/B_{\phi} = 1/2$, the vortices
form an ordered checkerboard state of a single domain. This state was also
observed in imaging experiments on vortices in superconductors
\cite{Harada10,Field12,Bending13} as well as of vortices in wire networks
\cite{Runge36,Hallen37}. 
In the experiments by Field {\it et al.} \cite{Field12},
domains of two different orientations of the checkerboard state are found at 
$B/B_{\phi} = 1/2$ where the domains of a single orientation are very large. 
It is unclear whether these domain structures in experiments result from
some disorder at the pinning sites or are an intrinsic part of the vortex
configurations as suggested by simulations of Josephson-junction arrays
\cite{Jensen_9237a}. We have conducted larger simulations on systems with up
to 392 vortices at $B/B_{\phi} = 1/2$; however, we have found only single
domains of vortices. In Fig.~2(d) the $B/B_{\phi} = 1/3$ state is shown.
Here the vortex configuration is not completely ordered; however, the
vortices are generally positioned in diagonal stripes running along
$\pm 45^{\circ}$ separated by two empty diagonal stripes. This state looks
almost identical to the results found in experiments \cite{Field12,Bending13}. 
Completely ordered states at the $1/3$ filling have been predicted for
repulsive particles on a square lattice as well as vortices in square
networks \cite{Halsey22,Lomdahl_9038}. The disorder in our system may come
from having too short of an annealing time. We have tested annealing times
from $2.5\times10^{5}$ to $5\times10^{6}$ normalized time units for
$B/B_{\phi} = 1/3$ and have not found any appreciable differences in the
vortex configurations; neither have we found noticeable system size
effects. We find that for weaker pinning, the diagonal ordering becomes
less prominent as some vortices begin to sit in the interstitial regions 
suggesting that, for our simulations and possibly for the experiments, the
vortex-vortex interaction, which tends to make the vortex lattice 
triangular, causes the disorder. In subsection B we show more explicitly 
the role of the pinning strength on the vortex lattice ordering at the
$1/3$ matching field. 

In Fig.~2(e)
at $B/B_{\phi} = 1/4$ the vortices form an ordered lattice with
vortices occupying every other pinning site in every other row.
In Fig.~2(f) we show the vortex lattice ordering at $1/5$ which shows
only regions of the square lattice ordering such as observed in the 
upper right area of Fig.~2(e). A similar ordering 
was observed in \cite{Field12}. 

\subsection{$V(I)$ Characteristics}
We show, in Fig. 3(a), a series of $V(I)$ curves for $B/B_{\phi} = 0.98$,
$1$, $1.05$, and $1.2$. The $V(I)$ curve shows a single jump at the matching
field. This is due to all the vortices depinning simultaneously. For
$B/B_{\phi} = 0.98$ we observe a two-stage depinning process with initial
depinning due to the onset of vacancy motion in the $B/B_{\phi} = 1$ 
commensurate vortex lattice, followed by depinning of the vortices at
the pinning sites which is associated with the second jump in the vortex
velocities. At $B/B_{\phi} = 1.05$ the $V(I)$ curves look similar to those
obtained in previous simulations for vortices in bulk superconductors
\cite{Reichhardt39}. Here the pinned phase is considerably reduced compared
to the $1/1$ filling, and a two stage depinning process can be seen. The
initial response is due to depinning of the interstitial vortices, and the 
second larger jump occurs when all the vortices begin to move. The additional
jumps in the vortex velocities are due to transitions between different
plastic flow phases which are discussed in more detail in Ref.\ 
\cite{Reichhardt39}.
For $B/B_{\phi} = 1.2$ the two stage depinning features are still present
and in general, for all $B/B_{\phi} > 1$, we observe the two depinning
thresholds.

In Fig.~3(b) we show velocity versus driving force for $B/B_{\phi} = 0.525$,
$0.5$, and $0.485$. At $1/2$ there is a single jump in $V_{x}$ while at
$0.525$ and $0.485$ the depinning transition occurs in two steps as seen
in the initial jump in the $V(I)$ followed by another larger jump. We find
that two stage depinning occurs just above or below the matching fields
where two distinct species of vortices exist: the vortices forming the
commensurate structure at the matching fields, and the vacancies or
interstitial vortices in the commensurate vortex structure. These vacancies
and interstitials will have well defined depinning thresholds lower than
those of the vortices forming the commensurate structure. For increasing
or decreasing vortex density, relative to the $1/2$ matching field, the
overall vortex structure can change and the distinction between the two
species of vortices is lost. In this case the two stage depinning
transition disappears as seen in the $V(I)$ curves in Fig.~3(c) for
$B/B_{\phi} = 0.3$ and $0.41$. Here the two-stage depinning process is clearly
absent. 

\subsection{Dependence of Vortex Configurations on Pinning Strength} 
We have carried out a series of simulations at $B/B_{\phi} = 1$, $1/2$, $1/3$,
and $0.43$ with $f_{p}$ ranging from $0.1f_{0}$ to $2.0f_{0}$. The vortex
configurations at $B/B_{\phi} = 1$ and $1/2$, as seen in Fig.~1(a,b), are
stable for all values of $f_{p}$ we have examined. In principle, even at
these fillings, for small pinning strength, the elastic interactions of
the vortex lattice will dominate, and consequently the lattice will be
triangular. 

For $B = 1/3$ and for $f_{p} > 0.6f_{0}$ the vortex configuration is the
partially disordered phase with alternating diagonal rows filled as seen
in Fig.~2(d). For weaker $f_{p}$ a portion of the vortices become located
in the interstitial regions between the pinning sites. In Fig.~4(a) we show
the vortices and the pinning sites and in Fig.~4(b) we show the vortices only 
for $f_{p} < 0.5f_{0}$ where we observe an ordering when the vortex-vortex
interaction is strong enough to induce triangular ordering; however, the
vortex lattice can still take advantage of the periodicity of the pinning
sites by allowing every other vortex to be pinned. In the ordered vortex
lattice of Fig.~4(a), vortices fill every third pinning site in every other
pinning row, while in the adjacent rows the vortices are not located
in the pinning sites but instead are positioned in every third interstitial
location. This produces a more triangular ordering in the vortex lattice 
which is easier to see in Fig.~4(b). As the pinning strength, $f_{p}$, is
reduced for $B/B_{\phi} = 0.43$, we observe some vortices sitting at
interstitial positions; however, we only observe regions of triangular
ordering for the weakest pinning strength.

In Fig.~5 we plot the critical depinning force, $f_{p}^{c}$, versus pinning
strength, $f_{p}$, for $B/B_{\phi} = 1/2$ and $0.43$. Here, $f_{p}^{c}$
increases linearly with $f_{p}$ for the entire range of values considered,
indicating that the vortices are in the single vortex pinning regime.
Since the pinned vortex lattice is symmetric, this can be understood by
realizing that the vortex-vortex interactions cancel and the depinning
threshold will therefore be the pinning force strength, $f_{p}$. For
$B/B_{\phi} = 0.43$ the critical depinning force increases in a non linear
fashion for low $f_{p}$ and then becomes linear for high $f_{p}$.
At $B/B_{\phi} = 0.43$ the vortex lattice is disordered and the
vortex-vortex interactions do not cancel, resulting in $f_{p}^{c}$ always
being less than $f_{p}$. The critical depinning force for $B/B_{\phi} = 0.43$
is always less than that of $1/2$ filling with the maximum difference
occurring near $f_{p} = 0.3$.  

\section{Moving Vortex States For Square Pinning Array}
Figure~6 shows the vortex positions, pinning sites, and trajectories for
the strongly driven vortex lattice for $B/B_{\phi} = 1$ [Fig.~6(a,b)] 
and $1/2$ [Fig.~6(c,d)]. Here we show that the moving vortex structures
are the same as the pinned vortex structures seen in Fig.~2. The depinning
at $1/1$, $1/2$, and $1/4$ filling fractions occurs elastically with vortices
retaining the same neighbors they had at depinning. The vortices are also
seen to flow in 1D channels along the pinning sites.   

For incommensurate fields, $B/B_{\phi} > 0.35$, the initial depinning is
plastic with vortices wandering in both the $x$ and $y$ directions. 
A transition to an ordered moving vortex lattice is found at higher drives. 
In Fig.~7 we show the vortex flow patterns for increasing driving force 
at $B/B_{\phi} = 0.64$. Just above depinning at $f_{d} = 0.805$ [Fig.~7(a)],
the vortex flow is plastic with the vortices wandering in both the $x$
and $y$ directions. For stronger drives, $f_{d}/f_{0} = 0.85$ [Fig.~7(b)],
the vortices begin to flow in 1D channels along the pinning sites; however,
there is still considerable hopping from one channel to the next. For
$f_{d}/f_{0} = 0.9$ [Fig.~7(c)], the vortex lattice is almost completely 
reordered with only occasional vortex jumps between adjacent channels.
For $f_{d} = 1.0$ [Fig.~7(d)], the vortex lattice is completely reordered
and the vortices flow in strict 1D channels along the pinning sites. In
Fig.~8 we show the vortex positions for $f_{d}/f_{0} = 0.805$ [Fig.~8(a)],
where the vortex lattice is clearly disordered, and $f_{d}/f_{0} = 1.0$
[Fig.~8(b)], where the vortex lattice has a defect-free distorted square
ordering. The reordered vortex lattice can be frozen into the zero drive
state by abruptly shutting off the driving; however, if the drive is slowly
decreased the vortex lattice will re-enter the plastic flow phase and 
disorder so no hysteresis is observed in the $V(I)$ curves.   

For $B/B_{\phi} < 0.35$ we find that the moving vortex lattice does not
completely reorder. In Fig.~9(a) we show the vortex positions, pinning
sites, and trajectories for $B/B_{\phi} = 0.28$ which indicate that the
vortices are flowing elastically in 1D channels along the pinning 
rows, but the number of vortices can differ from channel to channel.
In Fig.~9(b) the moving vortex lattice can be seen to retain a
considerable amount of disorder. We have found that for weaker pinning
the field at which the vortices can completely reorder is reduced below
$B/B_{\phi} = 0.35$. 

\section{Fractional Vortex 
Matching and Dynamics For Triangular Pinning Arrays} 

In Fig.~10 we show the critical depinning force versus field for a
system with triangular pinning. Matching effects can be seen at 
$B/B_{\phi} = 1$, $6/7$, $2/3$, $1/3$, $1/4$, and 
$1/7$. No depinning peak is seen
for the $1/2$ matching nor for $B/B_{\phi} > 1$. In Fig.~11(a) we show
the matching fields at $B/B_{\phi} = 1$, where all the vortices are
pinned forming a triangular lattice. In Fig.~11(b), for $B/B_{\phi} = 2/3$,
we find a honeycomb vortex lattice with vacancies forming a triangular
lattice, which is the same as the vortex lattice configuration at the $1/3$
matching field [Fig.~11(d)]. In Fig.~11(c), at $B/B_{\phi} = 1/2$, no
particular vortex lattice ordering is observable. In Fig.~11(e), for
$B/B_{\phi} = 1/3$, the vortices form a triangular sublattice. In
Fig.~11(f), at $B/B_{\phi} = 1/4$, the vortices again form a triangular
lattice while in Fig.~11(e), at $B/B_{\phi} = 1/7$, the overall vortex
lattice is mostly triangular with some defects. The vortex lattice
configuration at $B/B_{\phi} = 1/7$ (not shown) has a triangular 
configuration with a vacancy lattice the same as the vortex lattice in
Fig.~11(e). 

\subsection{Vortex Pinning and Ordering for Weaker Pinning Strength} 

As was shown in Fig.~11(c), the vortex configuration at $B/B_{\phi} = 1/2$
is disordered. In Fig.~12(a) we show that, for a weaker pinning strength 
of $f_{p}/f_{0} = 0.3$, an ordered state, in which half of the vortices
are located in the interstitial regions between the pinning sites and
the other half are located at the pinning sites, is possible for
$B/B_{\phi} = 1/2$. This gives the overall vortex lattice a square
symmetry which is evident in Fig.~12(b). 

In Fig.~13 we show $f_{p}^{c}$ versus vortex density for $f_{p}/f_{0} = 0.3$
for a more limited field range of $0 < B/B_{\phi} < 0.6$. Here $f_{p}^{c}$
is considerably reduced from that of $f_{p}/f_{0} = 0.9$; however, the
commensurability peaks are stronger with a larger difference between
the commensurate peaks and the incommensurate regions. In addition to 
finding the same commensurability peaks observed at $B/B_{\phi} = 1/3$, $1/4$,
and $1/7$, we now clearly observe peaks at $1/6$, $1/9$, and $1/2$. Most
experiments in periodic pinning arrays have found that commensurability
effects are strongest very near $T_{c}$ where the pinning is the weakest.
For decreasing temperature, the overall pinning increases, but the clear
matching anomalies gradually vanish. One possibility (in the experiments)
is that the bulk pinning becomes relevant for lower $T$; however,
experiments without dots or holes, for the same temperature regions, find
that the critical current is much {\it lower} than in the system with the
pinning arrays, indicating the strong pinning is coming from the dots or holes.       
Our simulations suggest that the increasing pinning force of the dots
alone can wash out the matching anomalies even in the absence of bulk
pinning. Figure~8 shows that the relative difference between the $f_{p}^{c}$ 
for a commensurate and incommensurate fields is maximum for low $f_{p}$
and is constant for higher $f_{p}$. The relative height of the commensurate
peak can be expressed as
$$
\delta f_{p}^{c} =  
\frac{(f_{p}^{c}({\rm com}) - 
f_{p}^{c}({\rm incom}))}{f_{p}^{c}({\rm incom})}
=  \frac{d_{c}^{p}}{f_{p}^{c}({\rm incom})}  . 
$$
For large $f_{p}$, $d_{c}^{p}$ will be constant and $f_{p}^{c}({\rm incom})
\propto f_{p}$ so $\delta f_{p}^{c}$ will decrease as $\sim 1/f_{p}$. 

The peak at $1/2$ occurs due to the formation of the ordered vortex
lattice as seen in Fig.~12. An experimental test for the presence of this
phase would be the appearance of an anomaly at $B/B_{\phi} = 1/2$
in a triangular pinning array as $T$ approaches $T_{c}$.

\subsection{Vortex Ordering for $1 < B/B_{\phi} < 1.5$}

In Fig.~14 we show the vortex configuration and pinning sites at 
$B/B_{\phi} = 4/3$, where it might be expected that the interstitial vortex
lattice will resemble the vortex configuration at the $B/B_{\phi} = 1/3$
matching field. Fig.~14 shows that a number of pinning sites are actually
unoccupied with two vortices sitting adjacent to the pinning sites and
caged in by six adjacent vortices. This indicates that placing an interstitial
vortex between two vortices at the pinning sites is costly enough to depin
one of the vortices. These types of defects disrupt any kind of overall
interstitial vortex lattice ordering so that higher field fractional
matching effects cannot be observed. For stronger pinning sites all the
pinning sites will be occupied so fractional matching may be possible. 

The general $V(I)$ characteristics are similar to those found in the
square pinning arrays with a two step depinning process for $B/B_{\phi} > 1$
and within $0.05$ flux-density of the sub-matching fields. We also find
that the general features of the vortex dynamics and reordering are similar
to those found with the square pinning array. 

\section{Conclusion}
We have investigated fractional matching effects in 
thin-film superconductors with square and triangular arrays of 
pinning sites. For the square arrays we find matching effects
in the form of enhanced critical depinning forces at $B/B_{\phi} = 1$, 
$3/4$, $1/2$, $1/3$, $1/4$, and $1/5$. The observed vortex configurations
at these fields are in good agreement with recent imaging experiments.
The vortex configurations also depend on the strength of the pinning sites.
The vortex configuration at the $1/3$ matching field can have a transition
from a diagonally ordered lattice to a distorted triangular lattice,
where every other vortex is interstitially pinned rather than being located
in a pinning site. 

We also find that the critical depinning current scales linearly with
$f_{p}$. Since the vortex lattice at the commensurate fields is symmetric,
the vortex-vortex interactions cancel and the vortex lattice is always in the 
individual vortex pinning regime. For large $f_{p}$, the critical depinning
force for the incommensurate fields also scales linearly with deviations from 
linearity at low $f_{p}$ as vortex-vortex interactions become relevant.

Near $B/B_{\phi} = 1, 1/2$, and  $1/4$ the $V(I)$ characteristics show
a two-stage depinning which is due to the presence of two well defined
vortex species: the vortices of the commensurate lattice configurations,
and the interstitials or vacancies in that lattice, which have different
depinning thresholds. For $B/B_{\phi} > 1$, a two stage depinning in the
$V(I)$ curves is always observed. 

We find that the vortex lattice depins elastically at commensurate fields
where the vortex lattice is symmetric; e.g., at $B/B_{\phi} = 1$, $1/2$,
and $1/4$, while at the incommensurate fields the vortex lattice first
undergoes plastic flow where the vortices flow in a random manner,
and then at a higher drives reorders and flows elastically in well defined
1D channels along the pinning sites. For $B/B_{\phi} < 0.35$, vortices will
still flow elastically along the pinning sites; however, there is considerable
disorder in the vortex lattice. For weaker pinning the vortex lattice will
reorder for fields $B/B_{\phi} < 0.35$. 

We have also investigated triangular pinning arrays and find matching
effects at $B/B_{\phi} = 1$, $6/7$, $2/3$, $1/3$, $1/4$, $1/5,$ 
and $1/7$, which
show ordered vortex lattices, while $1/2$ and other incommensurate fillings
show disordered vortex lattices. For $B/B_{\phi} < 0.5$ and for weaker pinning,
we observe additional matching at $2/7$ and $1/6$ as well as a general
enhancement of the other commensurate peaks. For the weaker pinning we also
find a commensuration peak at $B/B_{\phi} = 1/2$, which occurs due to the
formation of an ordering state where half the vortices are located in the
interstitial regions so the overall vortex arrangement is square.
The presence of these ordered states should be observable in magnetization
curves. 

For triangular pinning arrays, with $B/B_{\phi} > 1$, we do not observe
any matching features due to certain pinning sites remaining vacant
where a pair of vortices can rotate around the vacant pinning site.
We observe similar $V(I)$ characteristics and dynamics for the triangular
pinning array as seen in the square arrays.

The static states for the pinning arrays should be observable with
Hall-probe arrays, Bitter decoration and Lorentz microscopy. Similar
techniques may also be able to observe the dynamics; particularly at
strong drives when the vortices should move in well defined 1D channels
along the pinning. The motion of interstitials or vacancies near
commensurate fields can be tested experimentally by looking for multiple
depinning stages in the current-voltage curves.   

\begin{center}
{\bf Acknowledgements}
\end{center}
We thank C.J.~Olson for critical reading of this manuscript and
S.~Bending, S.~Field, F.~Nori, C.J.~Olson, I.K.~Schuller, 
and G.T.~Zim{\' a}nyi for useful discussions. This work was supported by 
NSF-DMR-9985978, by the
Director, Office of Advanced Scientific Computing Research, Division of 
Mathematical, Information and Computational Sciences of the 
U.S.\ Department of Energy under contract number DE-AC03-76SF00098 as 
well as by CLC and CULAR (Los Alamos National Laboratory/ University of
California).

\begin{figure}
\centerline{
\epsfxsize=3.5in
\epsfbox{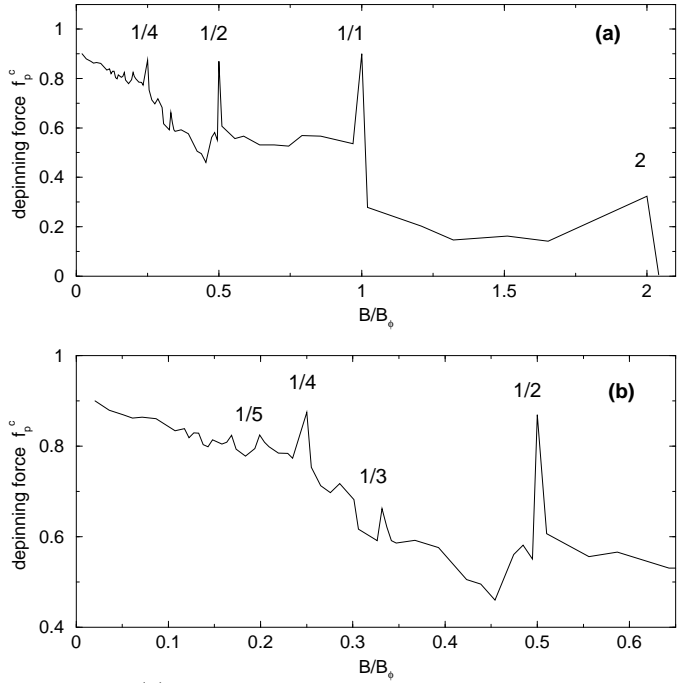}}
\caption{
(a) Critical depinning force $f_{p}^{c}$ versus vortex density 
$B/B_{\phi}$ for $ 0 < B/B_{\phi} < 2.1$ with a square pinning array.
Peaks can be seen in $f_{p}^{c}$ at $B/B_{\phi} = 1/4, 1/2, 1/1$, and $2/1$. 
(b) Critical depinning force for the system in (a) for $0 < B/B_{\phi} < 0.6$
showing narrowed view of the peaks in $f_{p}^{c}$ for $B/B_{\phi} = 1/5, 1/4,
1/3$, and $1/2$. Smaller peaks can also be seen at $B/B_{\phi} = 2/7, 1/6$,
and $1/8$.}
\label{fig1}
\end{figure} 
  
\begin{figure}
\centerline{
\epsfxsize=3.5in
\epsfbox{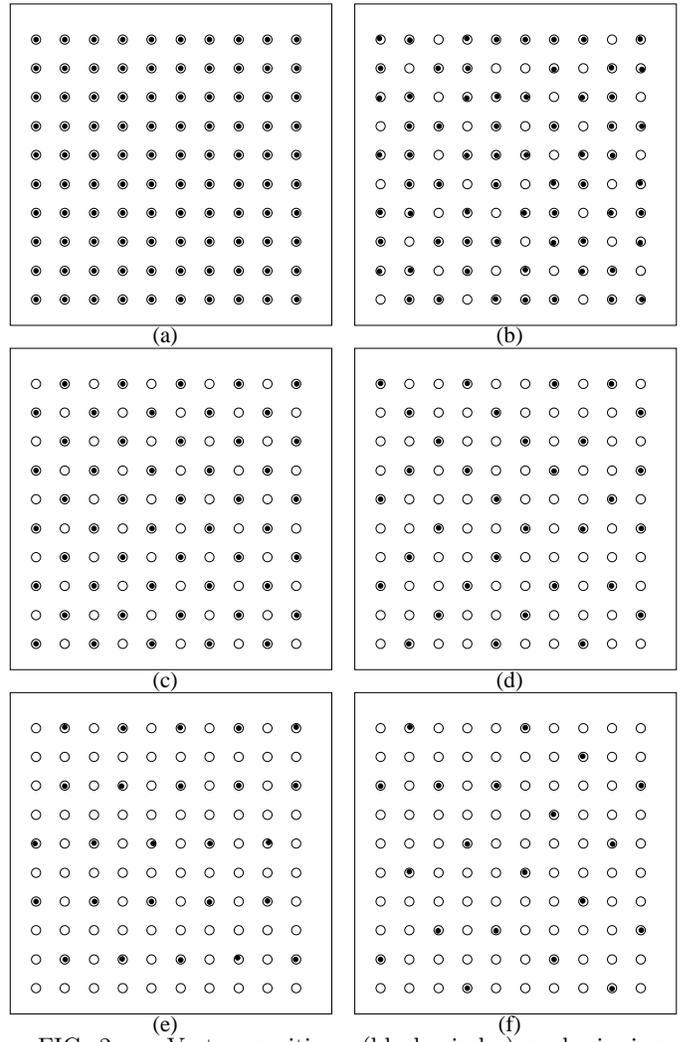}}
\caption{ 
Vortex positions (black circles) and pinning sites (open circles) from
the simulation in Fig.~1 for (a) 
$B/B_{\phi} = 1/1$, (b), $0.642$, 
(c) $1/2$, (d) $1/3$, (e) $1/4$, and (f) $1/5$.}
\label{fig2}
\end{figure}  
 
\begin{figure}
\centerline{
\epsfxsize=3.5in
\epsfbox{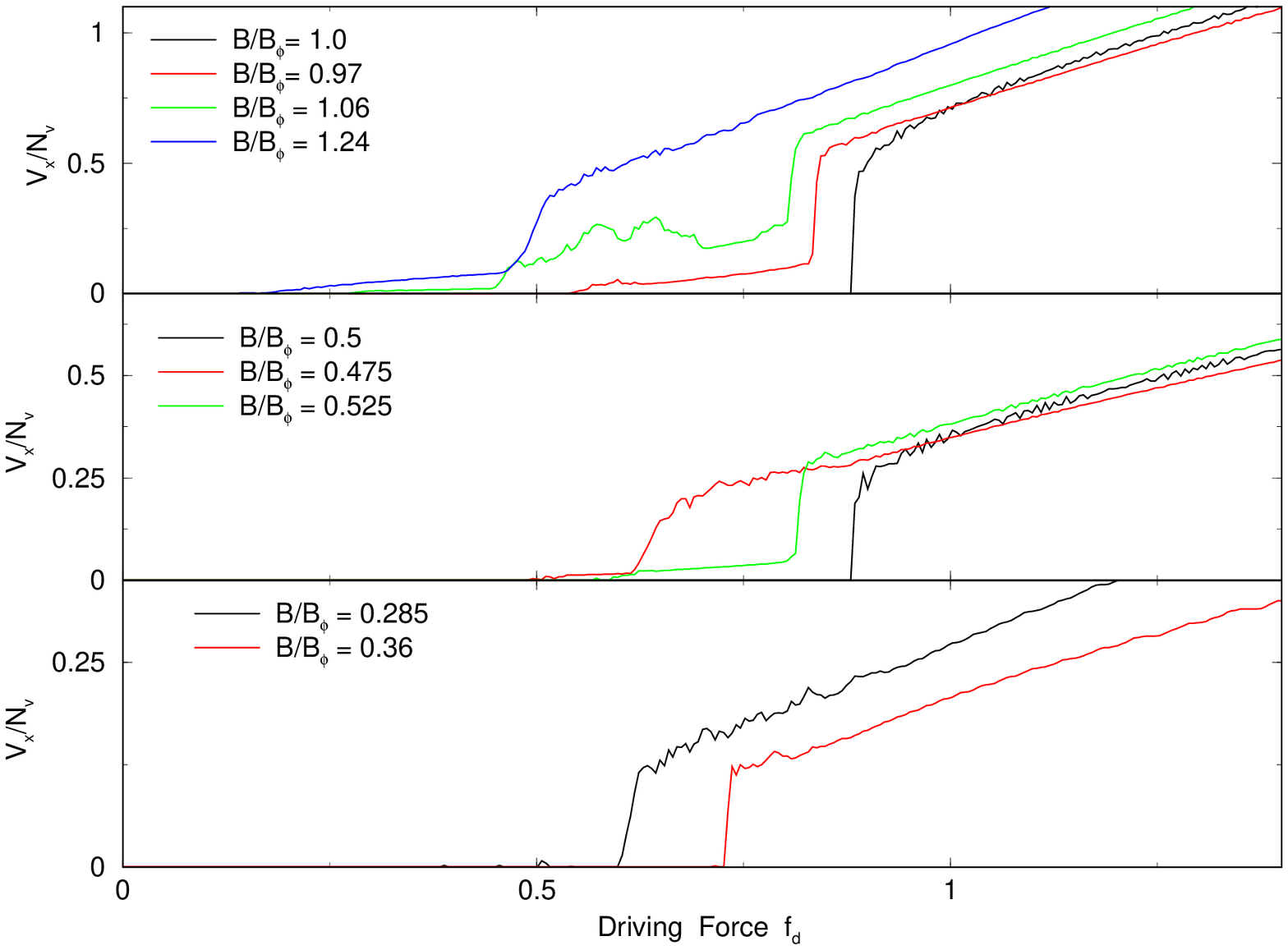}}
\caption{
Vortex velocities $V_{x}$ versus the applied driving force
$f_{d}$ for (a) $B/B_{\phi} = 0.98, 1.0, 1.06$, and $1.23$. The largest 
critical depinning force occurs for the 1/1 matching. (b) 
$B/B_{\phi} = 0.525$, $0.5$, and $0.485$. (c) $B/B_{\phi} = 0.30$ and $0.41$. 
}
\label{fig3}  
\end{figure}

\begin{figure}
\centerline{
\epsfxsize=3.5in
\epsfbox{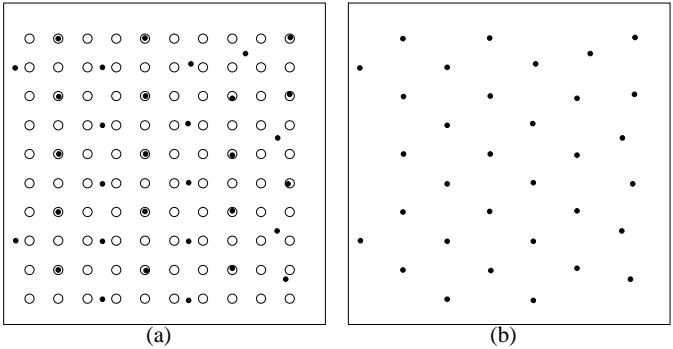}}
\caption{ 
(a) Vortex positions and pinning sites. (b) Vortex positions only for
$B/B_{\phi} = 1/3$ with $f_{p}/f_{0} = 0.3$.
Unlike the case for for $f_{p}/f_{0} = 0.9$ (Fig.~2(e)), where all the 
vortices are located at pinning sites, roughly half the vortices are
located at the interstitial regions between the pinning sites so that 
the overall vortex structure is a distorted square as seen in (b).}
\label{fig4} 
\end{figure}

\begin{figure}
\centerline{
\epsfxsize=3.5in
\epsfbox{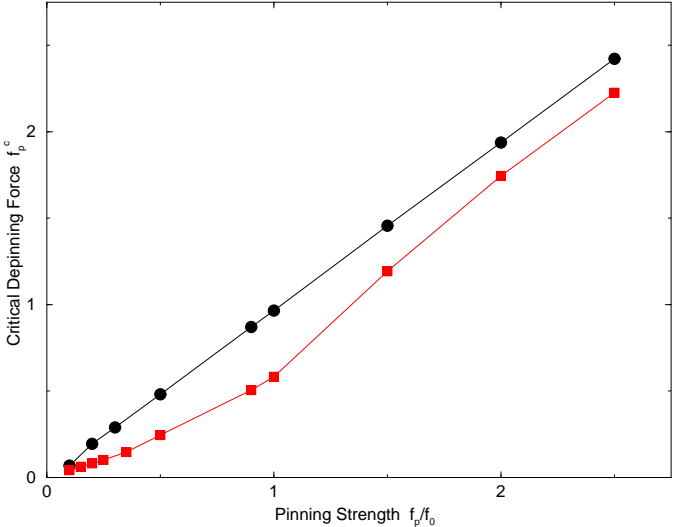}}
\caption{
Critical depinning force versus $f_{p}$ for $B/B_{\phi} = 1/2$ 
(upper curve) and $B/B_{\phi} = 0.642$ (lower curve). For $1/2$ filling
$f_{p}^{c}$ goes linearly with $f_{p}$ for the entire range of $f_{p}$
investigated. For $0.642$ filling $f_{p}^{c}$ is linear in 
$f_{p}$ for large values of $f_{p}$ but is non-linear for low $f_{p}$. 
A maximum difference between the two curves occurs near $f_{p}/f_{0} = 0.4$.}
\label{fig5} 
\end{figure}

\begin{figure}
\centerline{
\epsfxsize=3.5in
\epsfbox{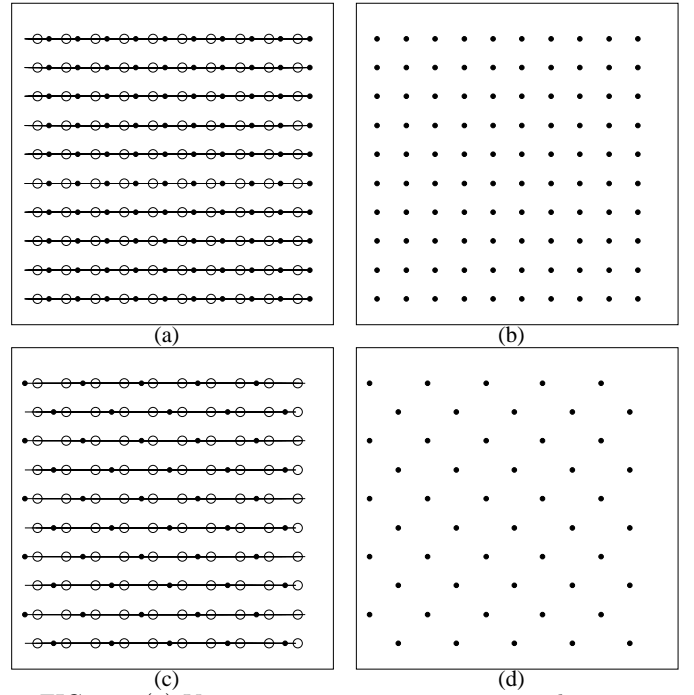}}
\caption{
(a) Vortex positions, pinning sites, and trajectories for the moving 
vortex lattice at $B/B_{\phi} = 1$. (b) Vortex positions showing the
moving square vortex lattice. (c) Vortex positions, pinning sites
and trajectories for the moving system at $B/B_{\phi} = 1/2$. (d)
Vortex positions showing the square moving vortex lattice rotated at 
$45^{\circ}$ with respect to the pinning lattice. In both cases the 
moving lattice has the same symmetry as the pinned lattice. Further,
the vortex motion can be seen to flow in well defined 1D channels 
along the pinning sites.}
\label{fig6} 
\end{figure}

\begin{figure}
\centerline{
\epsfxsize=1.75in
\epsfbox{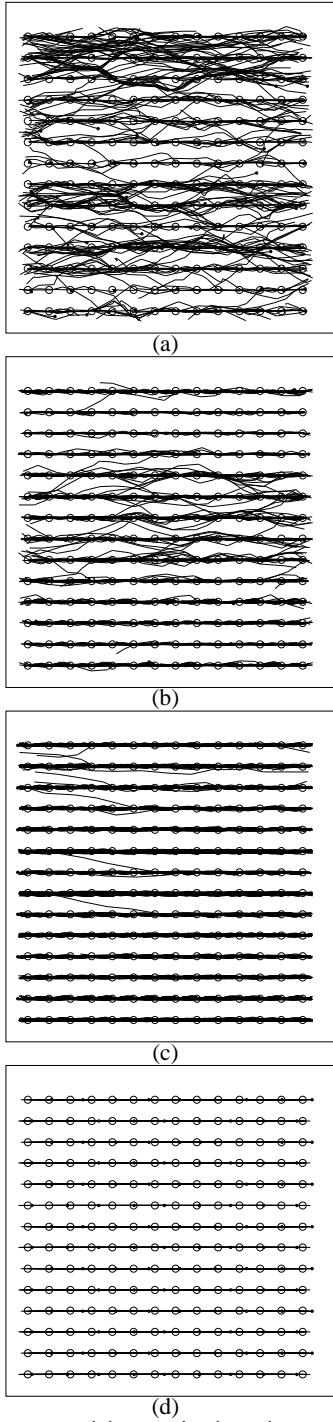}} 
\caption{
Vortex positions, pinning sites, and trajectories for $B/B_{\phi} = 0.6428$
for increasing applied driving force $f_{d}$. (a) For $f_{d}/f_{0} = 0.8$,
just above depinning, plastic flow occurs with vortices flowing in both $x$ and
$y$ directions. (b) At $f_{d}/f_{0} = 0.9$ the vortex motion is increasingly
along the pinning sites forming 1D channels. (c) At $f_{d}/f_{0} = 0.95$,
almost all vortex flow is along the pinning sites with only occasional
jumping of vortices between channels. (d) At $f_{d}/f_{0} = 1$, the vortex
lattice has completely reordered. Vortices flow plastically with the flow
restricted to 1D channels along the pinning sites. }
\label{fig7} 
\end{figure}

\begin{figure}
\centerline{
\epsfxsize=3.5in
\epsfbox{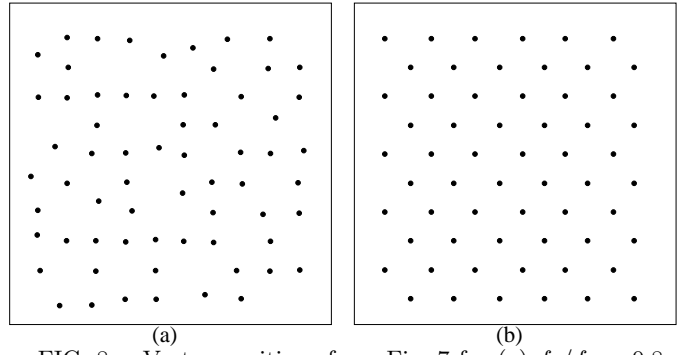}}
\caption{
Vortex positions from Fig.~7 for (a) $f_{d}/f_{0} = 0.8$ and (b)
$f_{d}/f_{0} = 1.0$. The strongly driven vortex lattice reorders to
a distorted square lattice.} 
\label{fig8}
\end{figure}

\begin{figure}
\centerline{
\epsfxsize=3.5in
\epsfbox{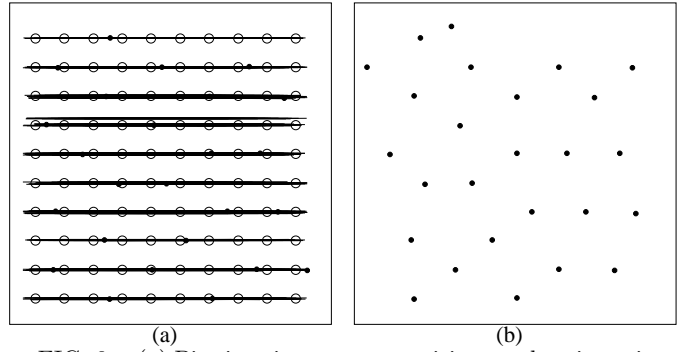}}
\caption{
(a) Pinning sites, vortex positions
and trajectories for $B/B_{\phi} = 0.28$ for $f_{d}/f_{0} = 1$. 
The vortices flow elastically in 1D channels; however, the number
of vortices is not equal in each channel.
(b) Snapshot of the vortex positions, showing that 
the moving vortex lattice retains a considerable amount of disorder. 
}
\label{fig9} 
\end{figure}

\begin{figure}
\centerline{
\epsfxsize=3.5in
\epsfbox{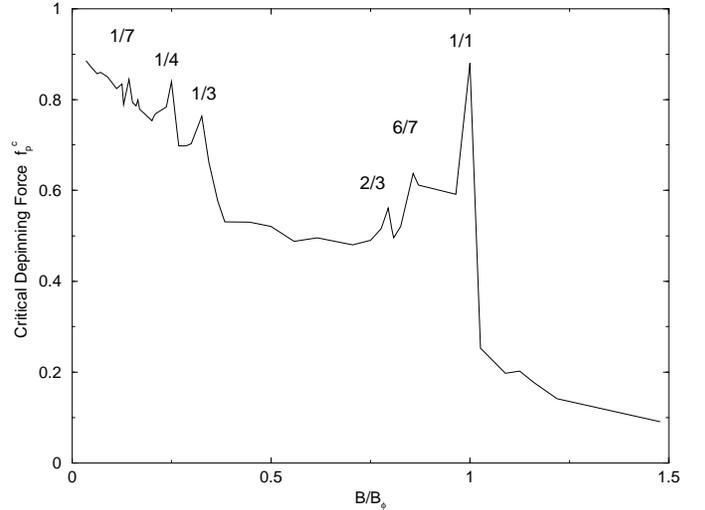}}
\caption{
Critical depinning force $f_{p}^{c}$ versus vortex density for a triangular
pinning array. Commensuration peaks can be seen at $B/B_{\phi} = 1/1$,
$6/7$, $2/3$, $1/3$, $1/4$, and $2/7$. Note that there is no peak at
$B/B_{\phi} = 1/2$.}
\label{fig10}
\end{figure}  

\begin{figure}
\centerline{
\epsfxsize=3.5in
\epsfbox{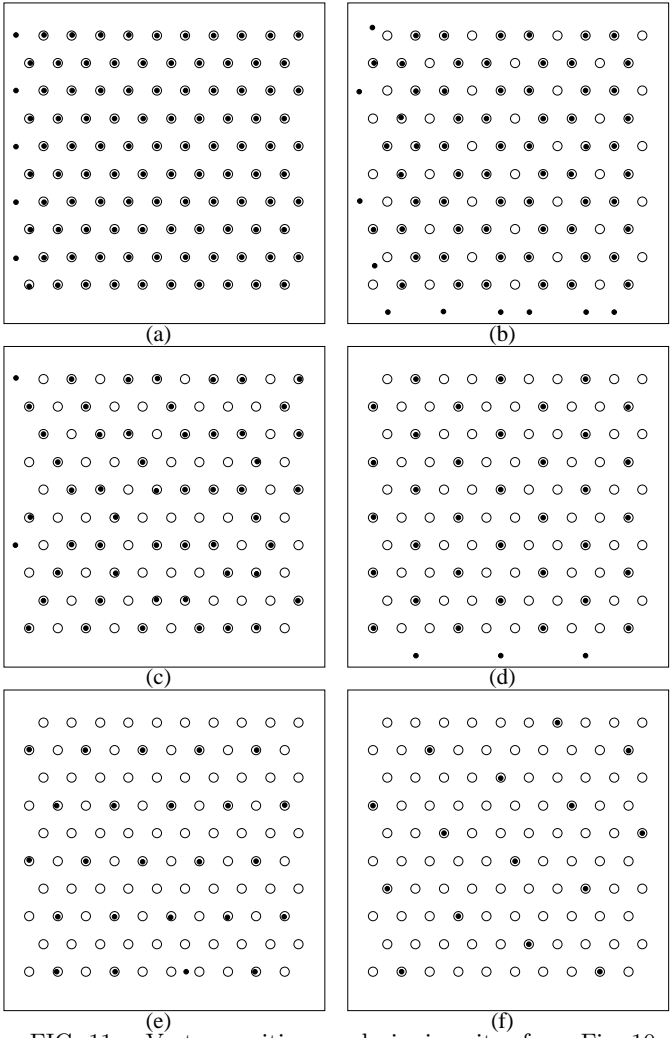}}
\caption{
Vortex positions and pinning sites from Fig.~10 for 
(a) $B/B_{\phi} = 1/1$, (b) $2/3$, (c) $1/2$, (d) $1/3$, 
(e) $1/4$, and (f) $1/7$.}
\label{fig11}   
\end{figure} 

\begin{figure}
\centerline{
\epsfxsize=3.5in
\epsfbox{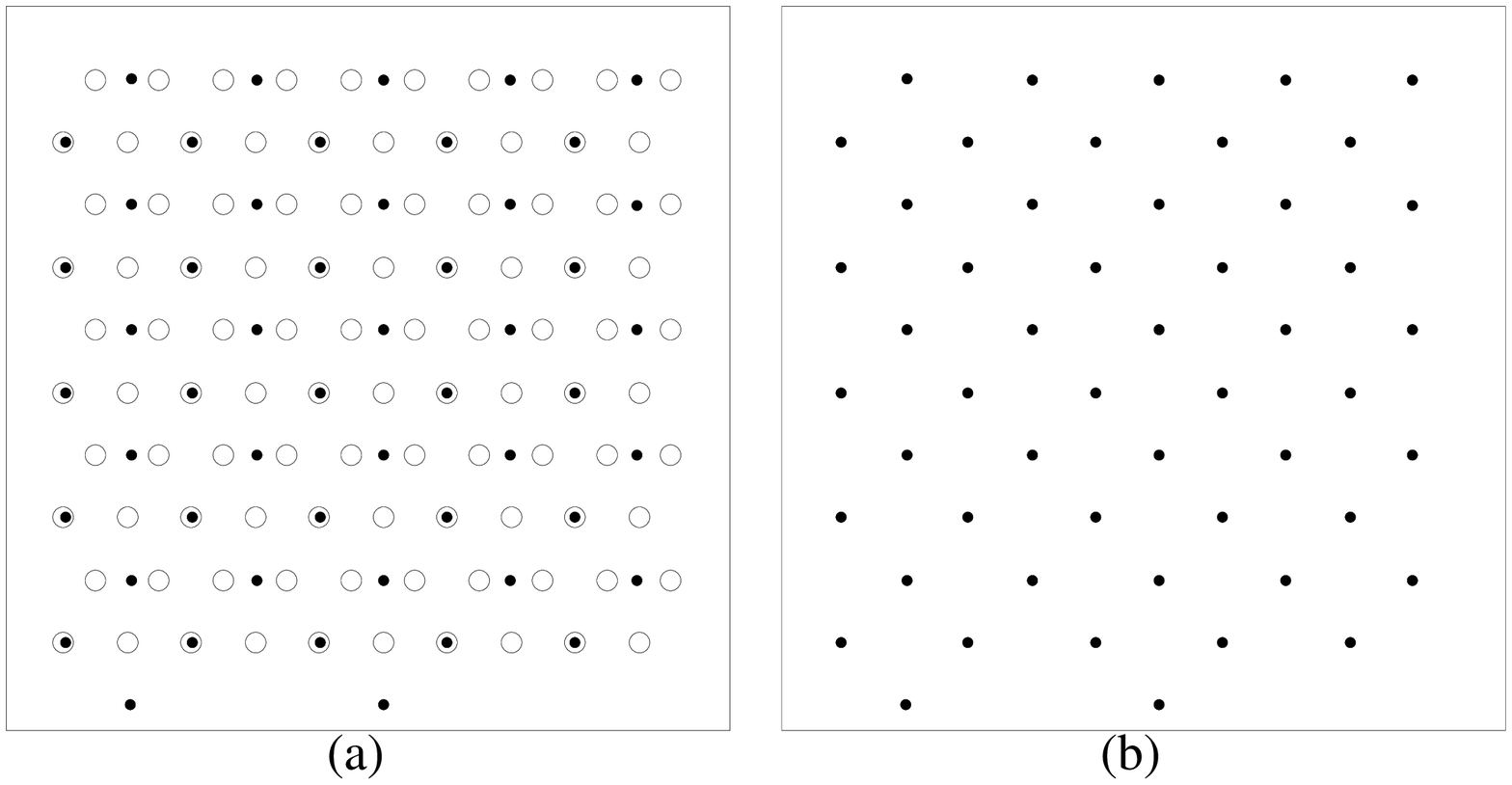}}
\caption{
(a) Vortex positions and pinning sites at $B/B_{\phi} = 1/2$ and (b)
vortex positions only for $f_{p}/f_{0} = 0.3$. In (a), only half of the
vortices are located at pinning sites with the other half located at
the interstitial regions. In (b) the overall vortex lattice has a square
symmetry.}
\label{fig12} 
\end{figure}

\begin{figure}
\centerline{
\epsfxsize=3.5in
\epsfbox{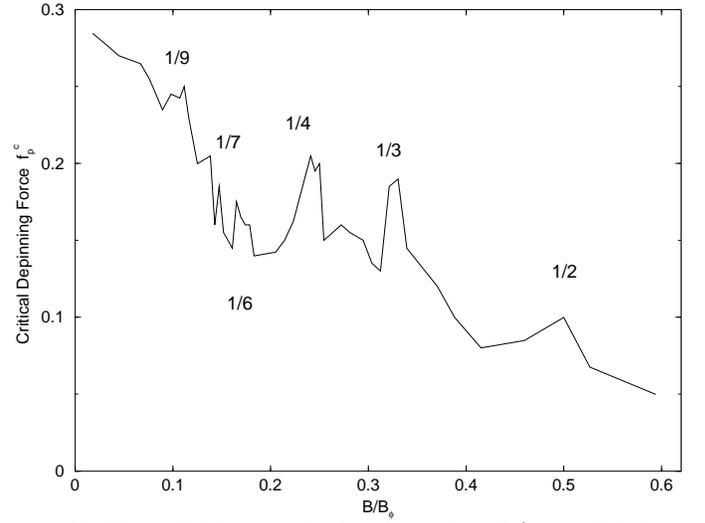}}
\caption{ 
Critical depinning force for $f_{p}^{c}/f_{0} = 0.3$ for
$0 < B/B_{\phi} < 0.6$. Commensurate peaks are observed at
$B/B_{\phi} = 1/4, 1/3, 1/6, 2/7, 1/7$, and $1/9$. A peak at
$B/B_{\phi} = 1/2$ is now visible with the vortices forming the ordered
state observed in Fig.~10.}   
\label{fig13}
\end{figure}

\begin{figure}
\centerline{
\epsfxsize=3.5in
\epsfbox{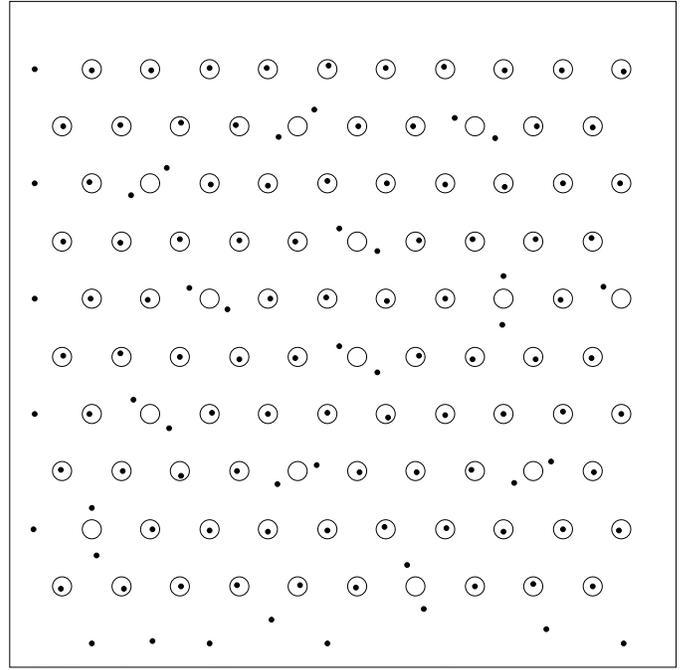}} 
\caption{
Vortex positions and pinning sites at $B/B_{\phi} = 1.25$ showing
that several of the pinning sites are unoccupied.}
\label{fig14}
\end{figure} 

\end{document}